\documentclass[a4paper,12pt]{article}
\pdfoutput=1
\usepackage{authblk}
\usepackage[utf8]{inputenc}
\usepackage[english]{babel}
\usepackage{graphicx}
\usepackage{amsmath}
\usepackage[pdftex]{hyperref}
\usepackage{url}
\usepackage{color}
\usepackage{natbib}
\usepackage{csquotes}
\usepackage[title]{appendix}
\usepackage[capitalize]{cleveref}

\setcitestyle{aysep={}}

\newcommand{\avg}[1]{{\left<#1\right>}}

\crefname{appsec}{Appendix}{Appendices}

\title{Stochastic Block Model Reveals the Map of Citation Patterns and Their Evolution in Time}

\author{Darko Hric\thanks{darko.hric@aalto.fi}}

\author{Kimmo Kaski\thanks{kimmo.kaski@aalto.fi}}

\author{Mikko Kivel\"a\thanks{mikko.kivela@aalto.fi}}
\affil{Department of Computer Science, Aalto University School of Science, P.O.  Box 12200, FI-00076, Finland}

\begin{document}
\maketitle

\begin{abstract}
In this study we map out the large-scale structure of citation networks of science journals and follow their evolution 
in time by using stochastic block models (SBMs).  The SBM fitting procedures are principled 
methods that can be used to find hierarchical grouping of journals into blocks that show similar incoming and outgoing citations patterns. These methods work directly on the 
citation network without the need to construct auxiliary networks based on similarity of nodes.  
We fit the SBMs to the networks of journals we have constructed from the data set of around 630 million 
citations and find a variety of different types of blocks, such as clusters, bridges, 
sources, and sinks.  In addition we use a recent generalization of SBMs to determine how much a 
manually curated classification of journals into subfields of science is related to the 
block structure of the journal network and how this relationship changes in time.  The SBM 
method tries to find a network of blocks that is the best high-level representation of 
the network of journals, and we illustrate how these block networks (at various levels of resolution) 
can be used as maps of science.
\end{abstract}

\section{Introduction}

The process of creating scientific knowledge relies on publications that are often stored and archived, with the primary purpose of preserving and distributing the knowledge obtained through research. These archives can also be used to study the science making itself, for example, by
extracting information of collaborations, citations, or keywords of the published articles.
Research in this field has a fairly long and rich history
with wide range of research topics, like the
assessment and prediction of performance and quality of individual papers, researchers, institutions, journals, fields, and even countries \citep{Taylor1967,Nerur2005,Lehmann2008,Althouse2009}, as well as 
identification of various large scale structures of science \citep{Price1965,Carpenter1973,Small1999,Waltman2010,Leydesdorff2013,Boyack2014},
journal classification \citep{Leydesdorff2006,Janssens2009,Zhang2010,Wang2016},
following research trends \citep{Porter2009,Persson2010,Chen2013},
and recognizing the emerging fields or researchers \citep{Small1989,Lambiotte2009,Cozzens2010,Shibata2011,Small2014}.

Bibliographic databases, like Web of Science, Scopus, and Google Scholar, store metadata of scientific publications,
which can be used to analyse science making at all levels, from large scale structure to performance of individual papers.
The number of \emph{entities} in the data, including articles, journals, citations, and scientists is very large and keeps growing exponentially (\cref{app:basic_stats}; \citealt{Pan2016arXiv}). To make sense of such massive amounts of data available about science one needs to simplify it and find its inherent patterns. This idea is not different from creating \emph{maps} that provide a simplified description of reality, i.e. maps of science that describes the endeavour of science in a broad sense \citep{Small1999,Boyack2005,Chen2013}. Such a map needs to provide a reasonably accurate simplification of the  structures it is mapping, i.e. individual elements need to be grouped (or clustered) to preserve large-scale patterns, while obscuring small and unimportant details. 
However, this is not a trivial task, and finding an optimal simplification accurately and reliably is becoming even more challenging 
as the networks under study continue to grow.

Conventional data analysis tools, such as clustering or dimension reduction methods, can be used to simplify 
the data about the complex relationships between the data entities. Representing the entities as vectors of their features is a common and practical abstraction that allows the use of clustering methods in the space of features, in which the most similar entities are grouped based on the similarity of the used features.
These vectors can contain citation information between the entities, and one can define similarity measures, like bibliographic coupling, co-citation, distance between citation vectors (Euclidean, cosine, 
Jaccard, etc.), and correlation coefficients between the citation vectors or publication texts (abstracts, keywords, etc.)
\citep{Kessler1963,Small1973,Marshakova1973,Carpenter1973,Leydesdorff2012,Boyack2005,Janssens2009}.

The data of scientific progress can be analysed with a variety of methods once the data has been preprocessed. The dimensionality reduction techniques 
project the vectors into the most significant subspaces
revealing groups of correlated entities (multidimensional scaling, factor analysis, etc.) \citep{Small1999,Leydesdorff2013}.
Classical clustering techniques, e.g. hierarchical clustering and k-means,
operate on the full space of features, and provide groups of similar entities,
based on implicitly or explicitly defined similarity measure or distance
\citep{Punj1983,Modha2000,Boyack2005,Silva2013}.
The factor analysis applied separately to the citing and cited 
direction of the complete 
citation matrix, enables further specialization into the types of clusters it finds,
since by using only one direction at a time, it detects clusters based on 
past and future citations, separately \citep{Leydesdorff2009}.
The co-citation and bibliographic coupling use similarities in citations in the future and past
respectively, and thus provide a separation naturally \citep{Weinberg1974}.
The results of this type of analysis depends on the prepossessing step of constructing the data vectors and similarities, and great care is needed in interpreting the results \citep{Boyack2005,Eck2009}.

The bibliometric data can also be analysed by constructing networks---such as the citation network between journals---and directly finding structure in them using the general purpose tools for analysing the networks.
The development of such methods within network science has exploded since massive amounts of data on large variety of networks---such as on social and transportation networks---have become available \citep{Newman2003,Boccaletti2006}. A prominent way of finding structure in citation networks using these methods is to investigate network clusters or communities \citep{Porter2009Communities,Fortunato2010,Fortunato2016Community}, which are subnetworks that have a large number of links inside them \citep{Rosvall2008,Lambiotte2009,Chen2010,Lancichinetti2012,Radicchi2012}. 
The assumption with most of these methods is that the network is constructed from densely connected cores of nodes or journals that have a relatively small number of citations to the rest of the network. This is in contrast to the methods based on similarity of journals that can find clusters with a strong preference for receiving or giving citations from a certain subset of journals, for instance work of applied research can cite theoretical works, without being cited back. 

Even if one would accept the premise that the community-like structures are relevant in citation networks, many community detection methods are besieged with intrinsic problems. Very often they detect structures even in case of random networks by mistaking noise for data, they might be very sensitive to small perturbations (noise), and posses a 
``resolution limit'', i.e. suffering from the inability to identify communities below a certain size that  depends on the total size of the network \citep{Guimera2004, Fortunato2007}. The performance, reliability, and even the results to some extent depend on the 
choice of a method from the large set of currently available methods.

The problems with community detection methods are well-known in the network science literature, and the need to find the richer structure in networks than 
those obtained by partitioning nodes to communities has been acknowledged for many types of networks \citep{Palla2005,Leskovec2009,Wang2010,Xie2013,Rombach2014Core}.
Very recently, as a solution to this problem, the old idea of using stochastic block models (SBMs) as models of network structure
\citep{Lorrain1971,Holland1983,Wasserman1987} has received renewed attention, because of the theoretical and algorithmic advances that enabled their use in a reliable and scalable way
\citep{Bianconi2009,Karrer2011,Peixoto2012}. 
SBM is a model in which nodes belong to \emph{blocks} and edges are created between (and within) the blocks with some fixed probabilities for each pairs of blocks. The methods based on SBMs work by finding the model which best explains the network data. 
The best explanation is not necessarily the model that would have most likely produced the data, but the simplicity of the model must also be taken into account, and 
the principled and powerful ideas from statistical inference literature are used to avoid such overfitting. 
One can consider the blocks as ``super nodes'' that are connected with weighted edges, and SBM methods then---by definition---try to find the ``super network'' that is the best simplification of the original network.

Here we take the advantage of the recent advances in SBM methods found in the network science literature and apply them to large scale citation networks between journals.
We use journal citation networks from Thomson Reuters Citation Index\textregistered~for the years ranging from 1900 to 2013 which contains hundreds of millions of citations. 
Unlike many previous in-depth studies of citation networks that have concentrated on small subsets of the citation network \citep{Pieters1999,Grossman2002,An2004,Nerur2005,Porter2009,Zhang2010,Shibata2011}, we 
focus on the large scale citation networks that are constructed using all articles in this bibliographic dataset. First we divide the full time period into the time windows of 5 or 10 years and use the articles in those windows to construct networks of the journals active in each window. That is, we take snapshots of the contemporary science at different points of time and track the important developments by fitting them with hierarchical SBMs. 
We visualize the resulting block structure at multiple levels of hierarchy, and illustrate the presence of blocks that are not community-like by categorizing them as sources, sinks, bridges, and communities. Moreover, we follow the evolution of these block categories in 16 largest fields of science in time and report  the 
large-scale changes in them over more than a hundred-year observation period.

The citation networks can be studied in isolation but they can also be augmented and compared with many other data sources such as journal categorizations, article keywords, and author information. Previous studies have, for example, compared predetermined journal categories to network clusters \citep{Boyack2005, Janssens2009} or to factors from factor analysis \citep{Leydesdorff2009}. They have also constructed networks using categories as nodes \citep{Zhang2010} and evaluated the quality of categorisations using criteria that favour cluster-like categories \citep{Wang2016}. Here we will utilize a recently developed generalization of the SBM method that allows the inclusion of any ``tag'' information about the nodes \citep{Hric2016} and use it to analyse how much information the predetermined journal categorizations carry about the block structure of the citation networks. This approach does not assume that the journal classifications are the ground truth, but determines the suitability of subject categories for describing citation structure by asking how much better we can do in estimating the citation flows with the classifications than we can do without the knowledge of the classifications. The construction of contemporary citation networks allows us to track the congruity of the subject categories with citation patterns throughout the last century.

The paper is organized as follows. The process of building annotated journal networks from raw 
citation data is described in \cref{sec:data}.
The stochastic block models are introduced and described in \cref{sec:SBM}. Then the 
visualization of the citation networks is described and a selection of results is presented in \cref{sec:vis_cit_flows}.
More detailed analysis of journal blocks properties is done in \cref{sec:conn_patt},
while \cref{sec:evolution} deals with their evolution in time. Next a comparison between the subject categories and citation structure is developed and presented in \cref{sec:pred_power_of_categories}.
Conclusions are made in \cref{sec:Conclusions}. 
Some basic properties of the data and additional results are presented in \cref{app:basic_stats,app:sf_predictiveness}.

\section{Data}
\label{sec:data}

All the networks constructed in this paper are based on data on articles and citations extracted from three \emph{Thomson Reuters Citation Index}\textsuperscript{\textregistered}~datasets 
(\emph{Science Citation Index Expanded}\texttrademark, \emph{Social Sciences Citation Index}\textsuperscript{\textregistered}, and \emph{Arts \& Humanities Citation Index}\textsuperscript{\textregistered}).
This database contains information about the publishing year and the venue  (journal, 
proceeding, conference, etc.) of articles, and each venue (from now on called \emph{journal}) is assigned 
to none, one, or several  \emph{subfields}. We join the subfields into larger \emph{fields} similar to \citet{Parolo2015}. The data set spans from  year 1900 to 2013 and contains $\sim$76 000 journals, $\sim$5.5M articles, and $\sim$630M citations in total. A more detailed description of the data can be found in \cref{app:basic_stats}.

As the full data set spans for more than a hundred years, it includes information needed to track development of modern western science. We aim to investigate how the citation patterns have evolved during this time period and to that end we split the data into multiple time windows, 
each of which is then used to construct a contemporary network of journals. The total volume of publications and citations is growing exponentially in 
time (\citealt{Pan2016arXiv}), and because of this we set the time window length to ten years before 1970s and to five years afterwards.

\subsection{Network construction}
\label{sec:data_networks}

In each time window, a node corresponds to an \emph{active} journal that has publications in the given time period. 
The connections between the journals are constructed using outgoing citations from these journals such that there is a directed link from 
journal \emph{a} to journal \emph{b} if an article in journal \emph{a} cites an article in journal \emph{b}, and the weight of this link is taken to be the number of such citations. For each time window we only count the contemporary citations satisfying the following two criteria:
(1) the cited article is published in a journal that is active in the time window, and 
(2) the time difference between the citing and the cited article is shorter than the length of the window. 
This procedure ensures that all articles in the time window contribute equally (with their citations) to network links. 

We have also tested a different approach for selecting the contemporary citations where both the citing and the cited article were required to be within the time window.
The more strict filtering of contemporary citations brings imbalance to incoming and outgoing citations of articles depending whether they are published at the beginning or towards the end of the window: those at the beginning 
have larger pool of articles they can receive citations from than the pool they can cite, 
and the opposite for articles towards the end of the window. We replicated all the results reported in this article using the networks created with this method, and they were very similar to the ones reported here.

\subsection{Simplified networks}
\label{sec:data_simple}

All of our results with the exception of the ones described in \cref{sec:pred_power_of_categories} are obtained using the weighted and directed networks of journals described above. 
Although all methods used in this manuscript
are fully capable of dealing with non-simple networks,
due to current technical limitations in the implementation of the method,
in \cref{sec:pred_power_of_categories} we used simplified networks (undirected, unweighted, and without self-loops).
A naive method of discarding link directions and weights,  
and removing self-loops, leaves the networks very dense, and is a poor approximation because 
it regards all links equally important, irrespective of their direction or weight. A usual approach, is to set a global threshold on the link weights that keeps only the strongest links, or to use only the links that form a maximum spanning  tree \citep{Kruskal1956,Macdonald2005}. Both of these are global methods, meaning that the decision on whether a link will be kept or removed depends on the weight distribution and the structure of the full network. We use a local thresholding method, in which statistical significance of weights of links of every node are calculated based on a null model defined for each node separately \citep{Serrano2009}. All links receive a score based on this criteria which, by excluding links above a certain threshold, allows us to keep only the desired number of links. We have tested the range of thresholds and find that the results are robust against the change of the threshold value (see \cref{sec:pred_power_of_categories}).

\section{Stochastic Block Model}
\label{sec:SBM}

Networks and graphs can be measured and summarized at many levels of granularity, starting from global or macroscopic measures---such as the total number of links or diameter---to local
or microscopic measures such as node degree or the clustering coefficient \citep{Newman2010Introduction}.
Here we concentrate on describing networks in a mesoscopic scale that is between these two extremes. 
Network analysis methods that work at this level of granularity almost exclusively deal with sets of nodes and links
called \emph{communities} or depending on the field of research, \emph{clusters}, \emph{groups}, \emph{modules}, etc.  \citep{Wasserman1994,Boccaletti2006,Schaeffer2007,Fortunato2010}.
There is not a single, precise definition of community, but most often it is described as 
a set of nodes with more connections between them than to the rest of the network \citep{Porter2009Communities,Fortunato2010}.
The community paradigm assumes that a network can be described as a collection of tightly knit sets of nodes, which are loosely connected to each other.

Stochastic block model relaxes the assumption about the nature of constituent sets of nodes such that they only need to be equivalent in the way they connect to other blocks, which in effect 
allows for a description beyond the community structure, like bipartite, core-periphery, 
etc. \citep{Karrer2011,Barucca2016}.
SBM is a generative model, meaning that it assumes a model of the underlying structure
and prescribes a procedure for building networks that have this structure in common.
The model is defined by assigning all nodes to disjoint sets\footnote{The assumption about blocks being disjoint sets of nodes can be relaxed \citep{Peixoto_model_2014}.} called \emph{blocks} and setting the number of links between and within blocks. Obeying the above described constraints, a network is generated  by randomly placing links between nodes. An equivalent description is to set 
the probabilities for placing a link between any two blocks, but we used the link counts
following the approach laid out in \citet{Peixoto_hierarchical_2014}.

Nodes within blocks share the probabilities for links towards the nodes in other blocks but also including their own block. In journal citation networks this means that all the journals in a block have the same citation patterns to other blocks.
They can, for instance, receive most of their citations from one set of journals, and give them out to another set, 
or have higher than average probability to exchange citations with some blocks and lower-than-average probability
with other blocks. Two blocks could also have identical citation patterns to other blocks, but different number of internal citations.
All this tells us that this model groups nodes (journals) into classes by their role in the network, whichever those are.

Once the model is known, building networks with the prescribed block structure is straightforward. However, the more common situation is opposite to this: 
only a single realization of the model of the empirical network at hand is known, and parameters of the model that most likely produce this network, need to be inferred. Finding the most likely parameters is a highly non-trivial task, and many approaches to solve it have been used \citep{Wasserman1987}.
All approaches use an objective function, in one form or the other, that measures the
probability of the given parameters to be the ones that produced the observed network.
The problem with this naive approach is that the best fitting model will be too detailed and will reproduce the observed network with very high accuracy,
which goes against the purpose of the models in providing a good simplification of the reality.
The cause for this is the fact that the simple approach uses all available data, including noise,  for fitting the parameters.
In the extreme case a highly-detailed model ends up putting all the nodes in their own blocks, since this reproduces
the network perfectly. A simple solution to prevent this from happening is to introduce
the number of blocks as a constraint in the fitting procedure \citep{Karrer2011}.
This works fine in cases where the number of blocks is known, otherwise it needs to be inferred from data,
for instance by finding a balance between the model description length and its goodness-of-fit to data.
Some of the approaches in this direction are listed in \citet{Latouche2012Variational}.
In this work we are taking advantage of the recent progress with this topic that estimates the number of blocks by including the information necessary for describing the model parameters into the total 
information amount being minimized such that it penalizes a too lengthy description \citep{Peixoto_parsimonious_2013,Peixoto_model_2014}.

One of the main issue with the classical SBMs, in addition to overfitting, is that 
they assume Poisson-distributed node degrees, so any deviation from the expected structure is considered a feature of the data and the fitting algorithm tries to find
a model that would adequately describe it. Fitting this model with a network with a different degree distribution would yield blocks that represent classes of nodes based 
on their degrees, not just on which other nodes they connect to.
So, for instance, highly cited 
journals would be separated from journals publishing review articles, that presumably cite 
in large volumes but do not get cited as much. Also, journals that play a central role in 
their respective communities would be put together, even if communities might not be
otherwise related.
The \emph{degree-corrected} stochastic block model is ``blind'' to this kind of structure:
it separates nodes into classes based on their degree, which removes unfair competition for
links between nodes of largely different degrees \citep{Karrer2011}.
This effectively includes the degree sequence into the model that does increase the information needed to describe the model, but the benefit of a simpler model for describing 
the data lowers the total description length in most cases.

The SBM and its degree-corrected variant can be expanded to describe hierarchical structure of 
blocks \citep{Peixoto_hierarchical_2014}. In this case, each level constitutes a network of blocks and the best fit of SBM of the level below it, starting from the network itself at the bottom level, to the 
trivial one-block level on top. Fitting all the levels is done simultaneously to obtain the minimum description length of the whole structure. This approach shows several advantages, like 
allowing for even smaller blocks that can reliably be inferred
and providing a multi-resolution view of the network.
Hierarchical SBM can be viewed as a stack of progressively simpler weighted 
networks---each level in the hierarchy is a zoomed-out version of the level below it.
This means that we can inspect connectivity patterns of groups of nodes at the desired level of detail/group sizes. 

The existence of the smallest identifiable blocks is in community detection literature
known as the \emph{resolution limit} and it is defined as the smallest group that 
a method is able to identify \citep{Fortunato2007}. One of the most commonly used objective functions is 
modularity \citep{Newman2004}, for which it has been shown that it is not 
able to separate small groups, even in obvious cases, if the number of links inside 
them is too small compared to the number of links in the rest of the network \citep{Fortunato2007}.
It should be noted that SBMs are not completely immune to resolution limit problems \citep{Choi2012,Peixoto_parsimonious_2013}, 
which presents a problem in the comparative analysis of networks spanning
two orders of magnitude, as is the case with our time slices. The smallest detectable 
blocks scale with the network size, which means that high-resolution levels for large 
networks would remain hidden from us.
Fitting the blocks at all resolutions for the hierarchical SBM at the same time reduces the limit, because each lower level uses blocks from the level above as a 
constraint, so block inference at the lower levels is in effect done locally in each higher-level block
\citep{Peixoto_hierarchical_2014}.

The criteria and algorithms described above are implemented in \texttt{graph-tool} Python module,
which we use to do all SBM fitting in this work \citep{peixoto_graph-tool_2014}.

\section{Visualizing citation flows between blocks}
\label{sec:vis_cit_flows}

Hierarchical SBM levels can be visualized as networks that provide us with a multi-resolution map of the citation network. Network visualization is a powerful tool for visual inspection of complex network data, 
but its readability and thus usefulness depends on the level of details presented and 
the total size of the network. In addition, if the network is dense, i.e. average node degree 
is large, it is even harder to make a clear picture of it. 
Lower levels contain a lot of information in fine detail, but they are often 
vast and too dense for visualization to be readable. But not all weighted links are 
equally important if we are interested in large-scale patterns. This means we can depict 
only the most important ones, making the network significantly less dense. This is again 
done by the procedure outlined in \cref{sec:data}, 
which keeps only the links that form the backbone of the network \citep{Serrano2009}.

\begin{figure}
  \centering
  \includegraphics[]{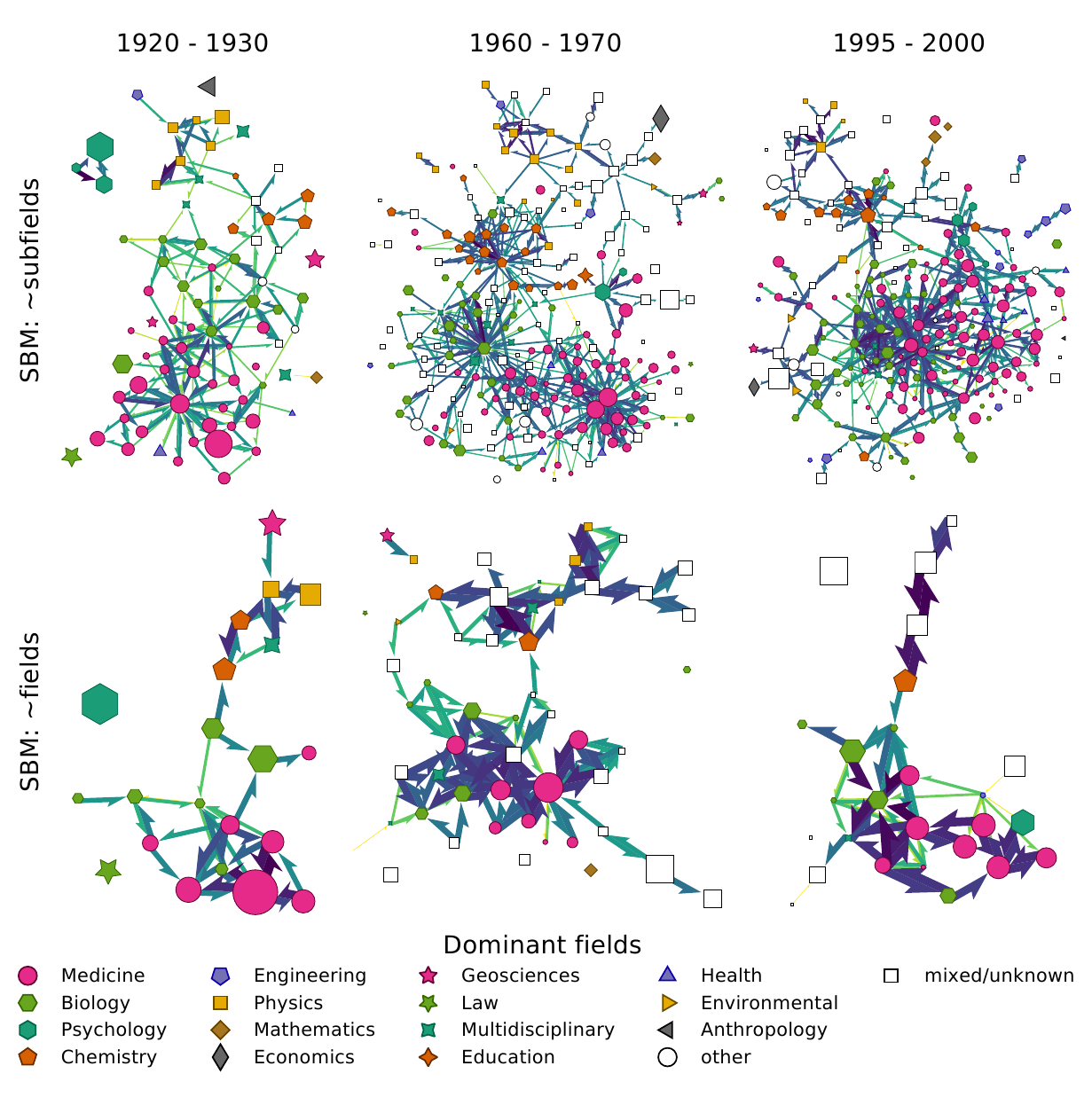}
  \caption{Networks of SBM blocks for the time periods of 1920s, 1960s, and 1995-2000. The top row 
shows resolution levels most similar to subfields, bottom to fields.
The node shapes and colours represent the dominant field in that block and 
the node sizes are proportional to the number of articles.
Directed links are coloured and sized logarithmically, according to the number of citations they carry.
Node and link sizes are normalized per pairs of networks from the same time window, so that these sizes
can be compared between the two resolution levels but not across the time windows.
The link colours are normalized for each network separately.
}
  \label{fig:networks}
\end{figure}

Three examples of networks of SBM blocks for the time periods of 1920s, 
1960s, and 1995-2000, at the levels most similar, i.e. the closest matching number of blocks to 
subfields and fields are depicted in \cref{fig:networks}.

Clustering of blocks of similar fields is quite visible in all six networks. Medicine 
forms the biggest cluster, followed by Biology, Chemistry, and Physics. Large-scale 
structure remains similar for all three time periods: Medicine is tightly 
connected to Biology, which is then connected to Chemistry and Physics.
This structure is in accordance with the previously published maps \citep{Rosvall2008,Leydesdorff2009}.
The figures also include interesting small-scale details that vary between the resolutions and time windows.
For instance, the interdisciplinary blocks are often located at the boundaries between other fields, but this effect is more visible in the subfield resolution level.

The fact that there are multiple blocks with the same dominant field in levels having 
the most similar number of blocks to the number of assigned fields, is a consequence of 
large heterogeneity of the field sizes and importance. Large fields also have rich internal 
structure that overshadows small, more simple fields in process of inferring the 
best blocks at each hierarchical level. In this sense, the choice of taking the number of 
fields as a guide for choosing the most appropriate level might be an overly simplified 
one.

\section{Connectivity patterns of blocks}
\label{sec:conn_patt}

In contrast to traditional community detection, blocks in SBM are not limited to ``dense 
subgroups'' where the citations stay inside the blocks, 
but a block can also represent a structure where the citations flow out of or into the block, as long as all journals in the block 
behave in similar way.
We summarize the type of block in terms of citation flows by counting the number of citations entering the 
block $s_{in}$ (articles in the block are being cited), leaving the block $s_{out}$ (they cite articles in other 
blocks) and internal citations $s_{int}$ (citing articles in the same block)\footnote{One
can view the system of flows and blocks as a weighted network.
In the literature of weighted complex networks \citep{Barrat2004}, weighted sum of a node's links is called \emph{strength}, and for the directed networks it 
can be in-, out- and internal strength: $s_{in}$, $s_{out}$ and $s_{int}$.}.

Note that the sum of the three flow counts represents the total activity of the journals in the block.
Here we are not interested in the total activities but in the type of flows.
We investigate these types by  separating the total flow from our flow measures and concentrate on relative in-, out-, and internal flows.
These normalized flows are defined as:
\begin{equation}
  \begin{aligned}
    \bar{s}_{in}=s_{in}/s_{tot}, \\
    \bar{s}_{out}=s_{out}/s_{tot}, \\
    \bar{s}_{int}=s_{int}/s_{tot}, \\
  \end{aligned}
  \qquad \text{where} \quad s_{tot}=s_{in}+s_{out}+s_{int}.
  \label{eq:s_def}
\end{equation}
By normalising the flows by the total activity we reduce the number of free parameters 
needed to describe blocks' flow patterns from three to two. That is, the sum of the 
three relative flows equals to one such that knowing two of them is enough as the third one can always be calculated based on them.
This means that we can report the two out of the three flow measures that are the most convenient for us. 

\begin{figure}
  \centering
  \includegraphics[width=.65\textwidth]{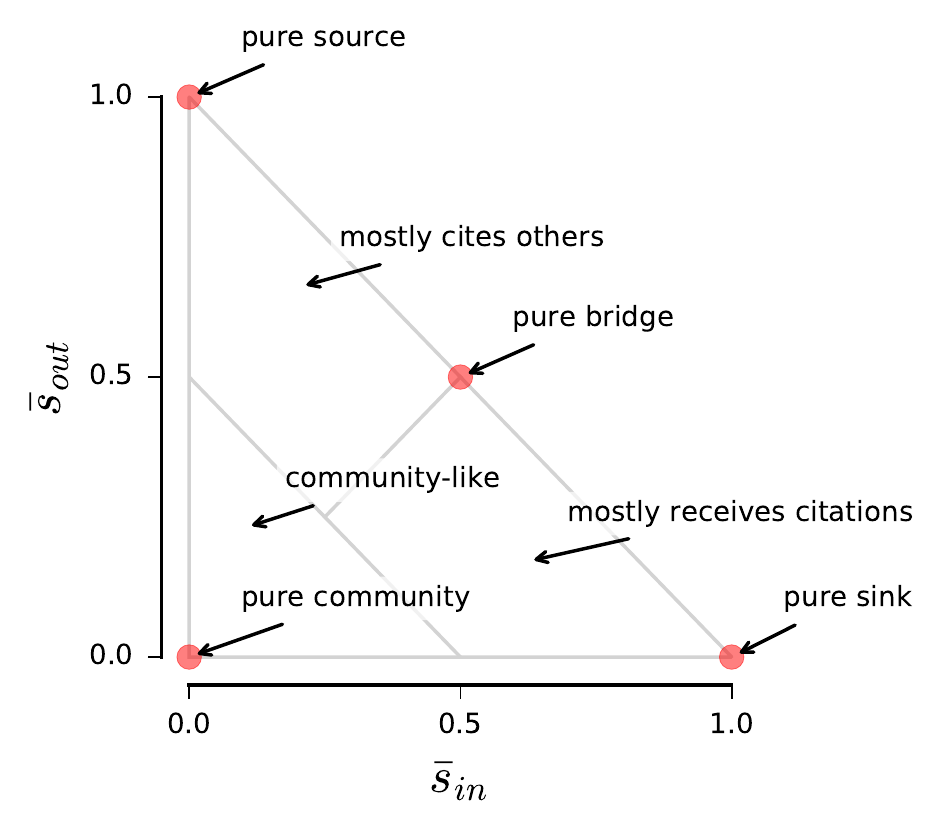}
  \caption{Properties of blocks depending on their location on the connectivity pattern 
plot. $\bar{s}_{in}$ and $\bar{s}_{out}$ are relative in- and out- flows of citations to/from the 
block.
Due to relation $\bar{s}_{in} + \bar{s}_{out} \leq 1$ points are confined to
regions outlined by grey lines.
Regions, as well as red points at special locations, are marked with arrows and
annotated with descriptions of the properties of blocks at those locations. }
  \label{fig:props_schema}
\end{figure}

For visualizing the connectivity patterns of blocks, the choice of $\bar{s}_{in}$ and 
$\bar{s}_{out}$ as $x$ and $y$ coordinates makes it easy to visually asses the block's 
properties from its position on the plot as illustrated in \cref{fig:props_schema}.
Since the sum $\bar{s}_{in}+\bar{s}_{out}$ must be $\leq1$, points can lie only in the
triangle bounded by the diagonal $(0,1)-(1,0)$.
Proximity to the origin 
tells us how much ``self-centred'' or ``community-like'' the block is, while the
distances from the axes signify the balance between receiving and giving out citations.
It helps to consider four extremal cases for a block, marked with red points in \cref{fig:props_schema}:
\begin{description}
  \item[\textbf{(0,0)}] \emph{Pure community}. Journals in this block are isolated from the rest of the network. Articles published in 
these journals only cite and get cited by articles in journals from this block.
  \item[\textbf{(1,0)}] \emph{Pure sink}. Journals in this block only receive citations and do not cite at all.
  \item[\textbf{(0,1)}] \emph{Pure source}. Journals in this block do not get cited, but cite others.
  \item[\textbf{(0.5,0.5)}] \emph{Pure bridge}. Journals in this block cite and get cited equally, but there are no citations within the block.
\end{description}
In most cases, the values lie somewhere between these extremes. 
The triangular space can be divided into three regions outlined by grey lines in
\cref{fig:props_schema} that contain blocks with the following properties:
\begin{description}
  \item[\textbf{Inner triangle}] is community-like (``a community in a weak sense'' [\citealt{Radicchi2004Defining}]). More than half of the citations pertaining to this 
block stay within it.
  \item[\textbf{Upper wing}] mostly cites others.
  \item[\textbf{Lower wing}] mostly receives citations.
\end{description}

\begin{figure}
  \centering
  \vspace{-80pt}
  \includegraphics[width=\textwidth]{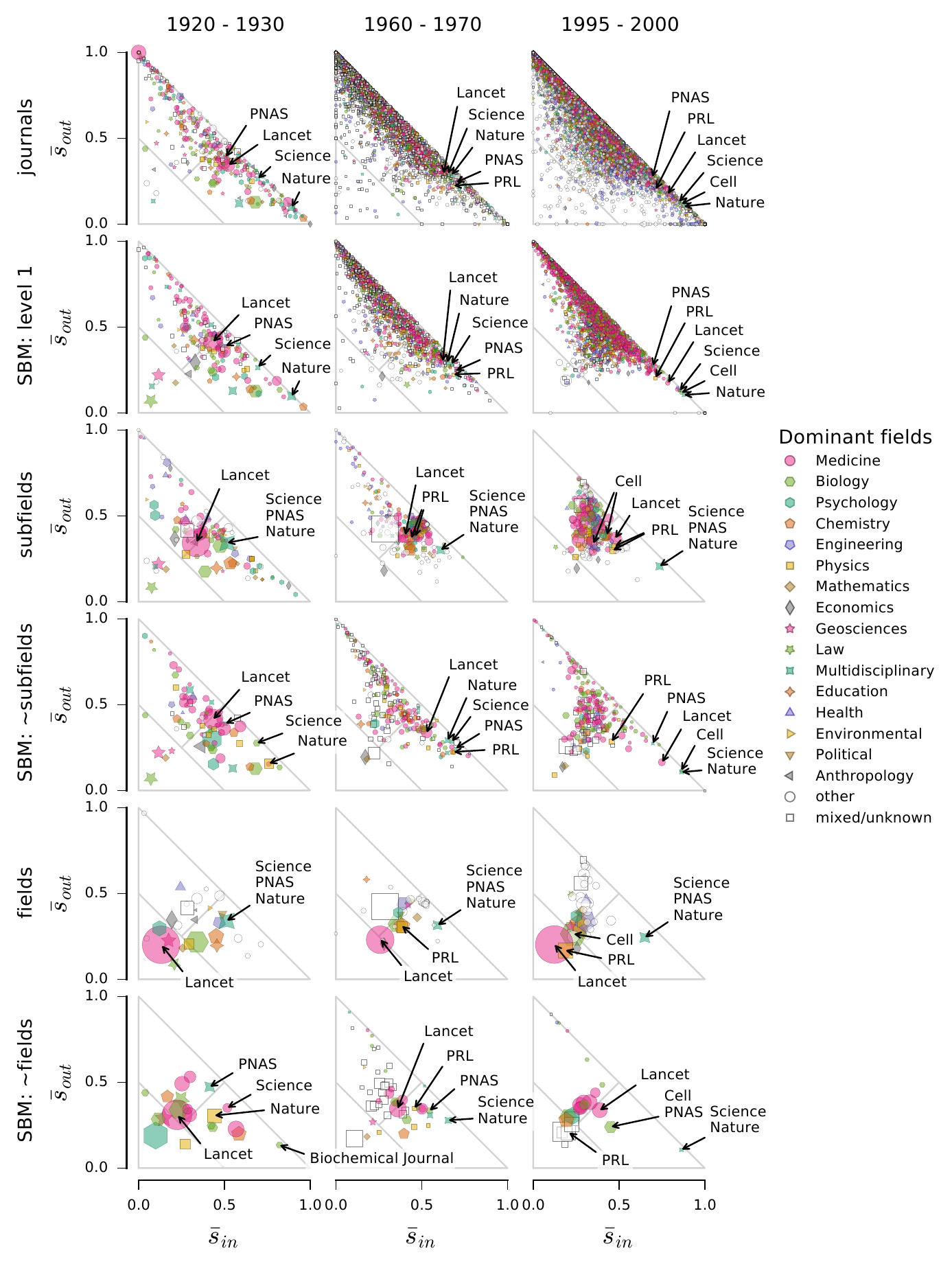}
  \caption{\small Connectivity patterns of structural and classification blocks, for the time periods of 1900--1910,
1950--1960, and 1995--2000. 
Each row corresponds to a different block type:
journals themselves;
the first SBM level;
subfields classification;
SBM level with the closest matching number of blocks to the number of subfields (SBM: 
$\sim$subfields);
fields classification;
and SBM level with the closest matching number of blocks to the number of fields 
(SBM: $\sim$fields).
Blocks are represented as points (coloured and shaped according to the dominant field in 
a block) with coordinates being the ingoing and outgoing citations as fractions of
total citations for each block ($\bar{s}_{in}$ and $\bar{s}_{out}$, \cref{eq:s_def}).
Point areas are proportional to the number of articles published by all of block's 
journals.
Blocks with fewer that 10 total citations are not shown.
Blocks containing a selected set of journals are annotated.
}
  \label{fig:block_props}
\end{figure}

In \cref{fig:block_props} we visualize the types of connectivity patterns of blocks found in the citation networks.
We display the data for three time periods (1920s, 1960s, and 1995-2000) and for three
levels of resolution. This gives us an overview of the types of mesoscale structures
one can find in the citation networks. A large number of blocks detected by the SBM method fall outside the inner triangle, and are thus not community-like structures even in the weak sense \citep{Radicchi2004Defining}. 
This is especially evident for high-resolution levels where the vast majority of inferred blocks are not communities.
Note that the inclination towards non-community-like structures is a feature of the data as the SBM method does not have a preference for any particular type of block structure.

Note that, in the hierarchical structure of blocks, the level that contains larger blocks cannot have less community-like blocks than a level that contains smaller blocks, and for the one block at the top of the hierarchy \emph{all} citations are internal.
This can be illustrated by considering the merger of two blocks: their internal citations remain internal, but part of their external citations that go between them become internal to the merged block (c.f. Appendix \ref{app:block_merge}).
In consequence, the average fraction of internal citations in the block at the higher level of hierarchy can only be equal to or higher than the weighted average of the two blocks at the lower level of hierarchy.

The outlined procedure can be used with any kind of partition of the journals to ``blocks'', not just ones inferred 
for SBM. 
The journals are explicitly partitioned in the data by subfield and field classification, and
we want to compare these partitionings to ones given by the blocks found by the SBM method.
For our purposes it is useful to view both of these partitions as block structures,
but to avoid confusion
we name the blocks as determined by the classification data \emph{classification blocks}, 
and the ones inferred by SBM from the citation patterns' \emph{structural blocks}.
Journals themselves form elementary blocks, which can be viewed as 
the ``zeroth'' level of hierarchical SBM or any other hierarchical block structure.
These zeroth level blocks give us properties of individual journals.
Journals are assigned to subfields (in the dataset), which are in turn grouped into 
fields. One would expect this classification to be reflected in citation patterns, 
since it should group similar journals together. We are able to test this assumption by comparing the properties of artificial blocks, 
defined by subfields and fields, with blocks inferred from citations. Hierarchical SBM 
provides us with many levels at different resolutions, and in \cref{fig:block_props} 
we compare the level
with the most similar number of blocks to the number of subfields and fields in that network.

Individual journals span the whole space of in- and out- flow balance (there are many
strong sinks as well as strong sources),
while the overwhelming majority do not predominantly 
cite themselves. There are annotated journals of higher prominence that are found in the lower wing, which is to be expected since
they receive more citations than give out.

The first SBM level determines the smallest non-trivial structural blocks,
and it is the smallest grouping of journals that cite, and are cited, 
in a similar way as they have similar citation patterns towards the rest of the network. 
Compared with the level of journals the spread of points is reduced in all time periods, 
although to a varying degree. The spread is reduced with time: there is almost no 
change in 1920s, some change in 1960s and a significant constriction of values for 1995-2000. This could be a consequence of the resolution limit, where smaller details are
increasingly harder to capture as the size of the network increases, or the outlying journals
carry less information in later years, so they are combined with more moderate ones.

At the level of subfields (the 3rd and 4th rows in \cref{fig:block_props}), both structural blocks and classification blocks display a pattern
where multidisciplinary blocks are located mostly on the cited side.
This is expected given that 
these blocks are dominated by high impact journals publishing high quality articles from a wide 
spectrum of fields. Comparing the distribution of points for classification blocks versus
structural blocks we see that the latter ones are more evenly spread out in 1960s 
and 1995-2000, while this is not the case in 1900s. 
There are also more community-like blocks 
in the structural case, in particular in the fields of Economics, Physics, and 
Mathematics and Geosciences in 1920s. This means that SBM captures a broader 
spectrum of block types, while classification blocks tend to be more similar to each other or
they have a preference for certain properties.

The highest level of hierarchy we focus on---for both classification blocks and structural blocks---is the level of fields
(5th and 6th rows in \cref{fig:block_props}).
Constriction in classification blocks is again present, albeit not so strong as for the 
subfields. With time, Multidisciplinary field strongly separates from the bulk that remains more elongated in the community-bridge direction and slightly leaned toward 
source-like behaviour. 
Similar to the level of subfields, Multidisciplinary block tends to become more cited with 
time for both classification and structural blocks, with this behaviour being more pronounced for 
structural blocks and the last time period.
Medicine and Physics get separated into a community-like block, meaning that the
most citations remain within their field, which was not the case for the subfields. The SBM finds
multiple blocks of mixed fields that are source- and community-like for 1960s, and only community-like for 1995-2000. Most of the mixed blocks in 1960s are 
comprised of unclassified journals, number of which rises in the second half of the century
(see \cref{app:basic_stats} for details). For classification blocks these journals are all collected under 
the ``mixed/unknown'' field (large white square).

Note that one needs to be careful when comparing blocks across different panels in \cref{fig:block_props}, as there 
is no guarantee that blocks with similar qualities in different panels comprise of the same journals. Deeper analysis in 
this direction would require listing all journals in a block, or annotating journals 
of interest (which is done for a few journals in \cref{fig:block_props}).

\section{Evolution of block connectivity patterns in time}
\label{sec:evolution}

In the previous section we summarized the types of citation flows of
individual blocks.
We will next build on this summarization method, and
quantify the evolution of citation flows of specific fields.
To be more precise, we ask the question:
what is the expected type of citation flow of the block where a randomly chosen article of a given field belongs to?

The expected flows for a journal in a field can be calculated 
by collecting all journals belonging to a field and taking a weighted average of the flows of the blocks they belong to.
To calculate average flows for articles, averaging needs to be weighted by a
fraction of journals' publications out of the total number in the field:
\begin{equation}
  \bar{s}^{f}_x = \frac{\sum_{j} a_j \bar{s}_x}{\sum_{j}a_j}\,,
\label{eq:sf_def}
\end{equation}
where $\bar{s}_x$ is any of the three average flows (in-, out-, and internal) from 
\cref{eq:s_def}, $a_j$ is the number of articles published by the journal $j$, and 
the sums go over all journals of the field $f$. If we do this for networks in all time windows, in 
addition to differentiating fields among themselves, we can follow the evolution of citation flow patterns of 
individual fields over time.

\begin{figure}
  \centering
  \includegraphics[width=\columnwidth]{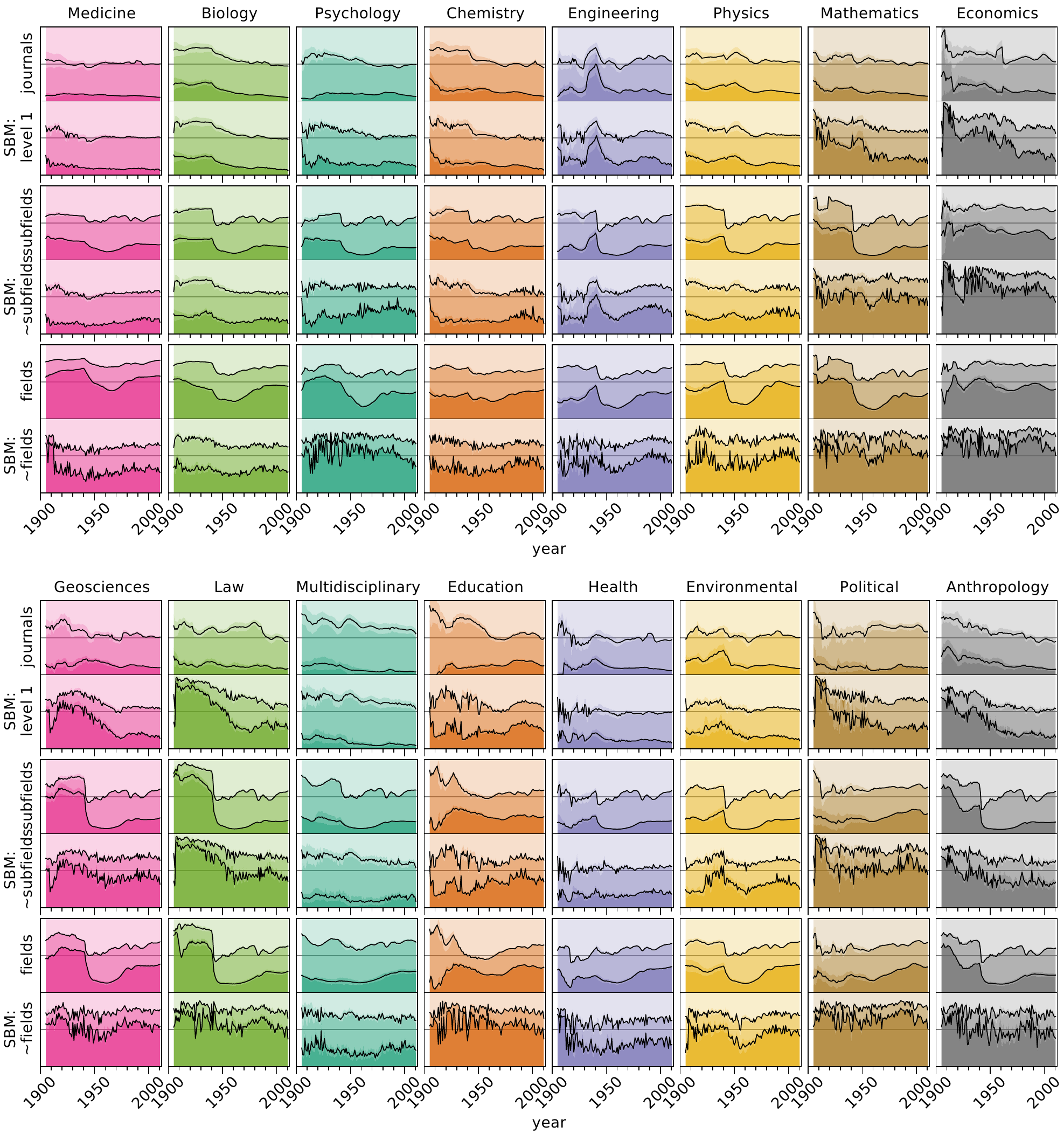}
  \caption{\small Evolution of structural and classification block connectivities in time, for 16 largest fields.
Fields are grouped in columns (colours are the same as in \protect\cref{fig:block_props})
and each row is for different type of blocks, from top:
journals themselves;
the first SBM level;
subfields classification;
SBM level with the closest matching number of blocks to the number of subfields (SBM: 
$\sim$subfields);
fields classification;
and SBM level with the closest matching number of blocks to the number of fields 
(SBM: $\sim$fields).
Time (in years) is on the horizontal axis.
Relative citation flows of blocks containing journals from the respective field 
[see \protect\cref{eq:sf_def}] are shown as shaded regions: bottom part (the darkest) is for 
internal flow $\bar{s}^{f}_{int}$, middle (lighter) for incoming flow $\bar{s}^{f}_{in}$, and top part (the lightest) for outgoing 
flow $\bar{s}^{f}_{out}$. The average value of each flow is marked with a black line, and the error of the mean is shown as 
a transitional shade between regions. The central horizontal line marks 50\% of the flow.
}
  \label{fig:block_props_in_time}
\end{figure}

The evolution of the average citation flows or block connectivities for 16 most prominent fields
is shown in \cref{fig:block_props_in_time}.
Fields are ordered by the surface area under the logarithm of the number of journals in each field.
Logarithmic scale is used to leverage the impact of the exponential increase of the number of journals (c.f. \cref{app:basic_stats}).
Similar to the previous section,
we repeat the analysis for structural blocks inferred for the SBM
and classification blocks given as classifications of the data (fields and subfields). That is, the average flows $\bar{s}^{f}_x$ (in-, out- and 
internal) of a field $f$ are calculated over all the field's journals, and the values $\bar{s}_x$ 
are taken from blocks,  either the structural blocks or the classification blocks, the journals belong to.

For half of the 16 fields, the internal flows of classification blocks are larger than the
internal flows of the inferred structural blocks.
Mathematics, Economics, Geosciences, Law, Education, Political Science, and Anthropology journals 
are found to reside in structural blocks that are more community-like than the corresponding classification blocks. This could mean that SBM is able to find their ``natural'' communities---ones 
that capture the most of the citation flows---while the classifications alone are not able to achieve.
The strength of this effect varies, with the most striking examples being Mathematics, Law, 
and Political Science. Classification blocks of Economics, and partially Geosciences and Law are 
themselves quite community-like.
The opposite effect is present for Medicine and Biology, meaning  that the structural blocks are less 
community-like than classification ones. 
This can be explained by the fact that these fields 
have a large number of both subfields and journals, so they contain a rich internal 
structure of blocks with large flow of citations between them.
Their internal citation structure might also be different from other fields, as they
contain highly cited papers describing methods and procedures \citep{Small1974}.

The in-flows $\bar{s}^{f}_{in}$ and out-flows $\bar{s}^{f}_{out}$ are quite balanced for most fields, with a few notable exceptions. 
Multidisciplinary field has noticeably more incoming citations than outgoing, for all hierarchy levels and for both types of blocks.
This does not necessarily mean that all journals in Multidisciplinary field attract 
citations, but it can be due to the fact that it contains several high profile
journals\footnote{For this reason some authors have in similar analyses excluded these journals altogether \citep{Zhang2010}},
like Nature, Science, and PNAS.
Individual journals in Political Science field receive more citations than they give out, 
while this difference vanishes for the blocks they belong to.
The opposite is true for Medicine, Health, and Environmental (they give out more citations than they receive) and this behaviour remains present for blocks in higher levels.

In time domain, fields exhibit a wealth of behaviours. Some have 
relatively stable patterns (Medicine, Chemistry, Economics, Multidisciplinary, and to some 
degree Health and Political Science), most of the others have large changes around the World War II,
while some have uniform and steady shifts (Education and Anthropology are the 
most notable).

The most striking feature is the sudden rise in share of outgoing flows at the 
time of the WWII, mostly for classification blocks, and to some degree for journals 
 of some fields. The largest rises are in decreasing order for: Law, Geosciences, 
Mathematics, Anthropology, Physics, Biology, and Environmental. In Biology and Chemistry a 
faint effect is also visible at the level of journals, while in case of Environmental it is mostly a 
drop in internal flow. Given that this effect is almost invisible for structural blocks,
the observed changes most likely do not arise from the change in journals' citation patterns,
but in the way they are classified.
This sudden change correlates with the large increase of the number of subfields (c.f.
\cref{app:basic_stats}).

Some fields exhibit slow but steady change over time, predominantly
in the structural blocks.
Anthropology journals are 
citing more external literature and less themselves with time, which is also  
visible in the smallest structural blocks, to lesser extent in the structural blocks of size of subfields,
and not at all in the structural blocks of size of fields.
This means that considerable amount of the citation flow that is increasingly going out of the journals is nevertheless retained
inside the structural blocks.
Biology, Chemistry, Multidisciplinary, and 
Education journals show similar behaviour, but their outgoing flows remain 
stable also in the structural blocks of size of subfields.
A plausible explanation is
that as a new journal appears in the field, it ``steals'' some of the citation flows 
from the old journals while mimicking their citation pattern towards the rest of the network,
which means that they will all be nevertheless put into the same block by SBM.
Proving such explanations would require a more detailed analysis.

\section{Predictive power of subject categorisations}
\label{sec:pred_power_of_categories}

Citation networks can be augmented with a wealth of information about articles, journals, and authors. 
These include subfields, tags, keywords, author affiliations, etc. 
In this work we use classification of journals into subject categories (subfields) provided in the dataset, and we are interested in how 
much does this classification correspond to the structural blocks found in the citation patterns.

The comparison of two partitions---such as classifications, clusterings, or block structures---is in general often 
done using some comparison measure, such as Jaccard index, 
Omega index, and Variation of Information \citep{Meila2007}.
It is typical to compare partitions arising from a classification given in the data and the groups arising from the network structure 
as returned by some community detection 
method \citep{Bommarito2010,Chen2010,Lancichinetti_comparison_2009,Hric2014,Yang2015}. This 
is a viable option in our case as well, but we would have to make a choice of a comparison measure.
Because the question of how similar two partitions are is ill-defined, each comparison measure
realises it differently and can even return different results \citep{Meila2007,Traud2011,Fortunato2016Community}.

Instead of asking how similar the two partitions are, 
we ask the question:
\emph{what can we know about the citations of a journal from its classification?}
Exactly this question is answered by including node annotations into SBM as it is done by \citet{Hric2016}, which
is based on the notion that annotations on nodes are just meta-information 
one has about the network---there is no principled difference between the data about 
connections between two nodes (links) and between a node and its annotations. In 
literature dealing with the community detection in networks this distinction
between data and annotations is often made explicit, 
either by treating annotations as a sort of ``ground truth'' for 
groups \citep{yang_community-affiliation_2012,yang_structure_2012,Yang2015},
or as features that need to be learned by the 
model \citep{Newman2016}. Here instead, annotations are treated as nodes of a bipartite 
network consisting of ``data'' nodes (journals in the citation network) and ``annotation'' 
nodes (subfields or fields of the journals), and connection exists between data node and all of its 
annotations (there can be any number of them, including zero).

\begin{figure}
  \centering
  \includegraphics[]{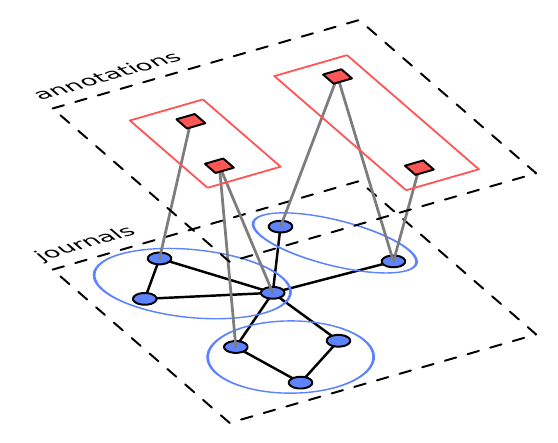}
  \caption{Schematic representation of joint journal-annotations model.
Journals with citations connecting them (blue circles and full lines) are augmented with
annotations (red squares and grey lines). SBM is fitted onto the whole network such that blocks of journals
are separate from blocks of annotation (blue circles and red squares respectively).
Note that a journal can have multiple annotations, or it can be unannotated.
Here we use subfield and field classifications as the annotations of journals, but any other data on
journals can be used.
}
  \label{fig:layers}
\end{figure}

\cref{fig:layers} illustrates the resulting combined network that consists of two kinds of nodes (journals and annotations)
and two kinds of links (citations and journal-annotation assignment).
SBM is then fitted with a constraint that each inferred block must contain only one kind of nodes.
The benefits of this procedure is that the node annotations contribute to the inferred node blocks,
and annotations are also grouped into blocks of ``equivalence''.

Working in this framework, the question from the beginning of this section
can be formulated as: 
\emph{how much information gain does one get about links of a single node, after learning the node's annotations?}
To answer it the following procedure is used. A small fraction of nodes is 
removed from the network (5\% or 100, whichever is smaller), turning them into 
``extra nodes''---the nodes we are missing the information on, and would like to know 
our chances in correctly guessing where their links connect to. Then, 
the blocks are inferred for both the original network (without annotations), and one
described above (with annotations included in an additional layer).
Without annotations, the only thing we know about extra nodes is their degree. How are 
these links distributed to existing blocks depends only on which block does the new node 
belong to (in SBM the only thing defining the node is its degree and block). 
However, we do not know which block the extra node belongs to, and its probability of being in a block can only be taken to be 
the size distribution of the blocks.
In case we have the node's annotations, we know its links in the annotations layer which 
narrows our choice of blocks it can belong to and thus raises the probabilities of 
guessing the correct links.
If we denote the probability for guessing all links of node $i$ without knowing 
annotations with $P_i$ and the same by using annotations as $P_i(ann)$ we can 
quantify the relative improvement with the \emph{predictive likelihood ratio} $\lambda_i$:
\begin{equation}
  \lambda_i = \frac{P_i(ann)}{P_i+P_i(ann)}\,.
\label{eq:lambda}
\end{equation}
The predictive likelihood ratio $\lambda_i$ takes values from $[0,1]$ and is above $0.5$ if annotations improve
link prediction power, around $0.5$ if they do not change it, and below $0.5$ if they 
decrease it. The average $\lambda_i$ of all sample nodes is the \emph{average predictive likelihood
ratio} $\langle\lambda\rangle$ for a dataset.

\begin{figure}
      \centering
      \includegraphics[]{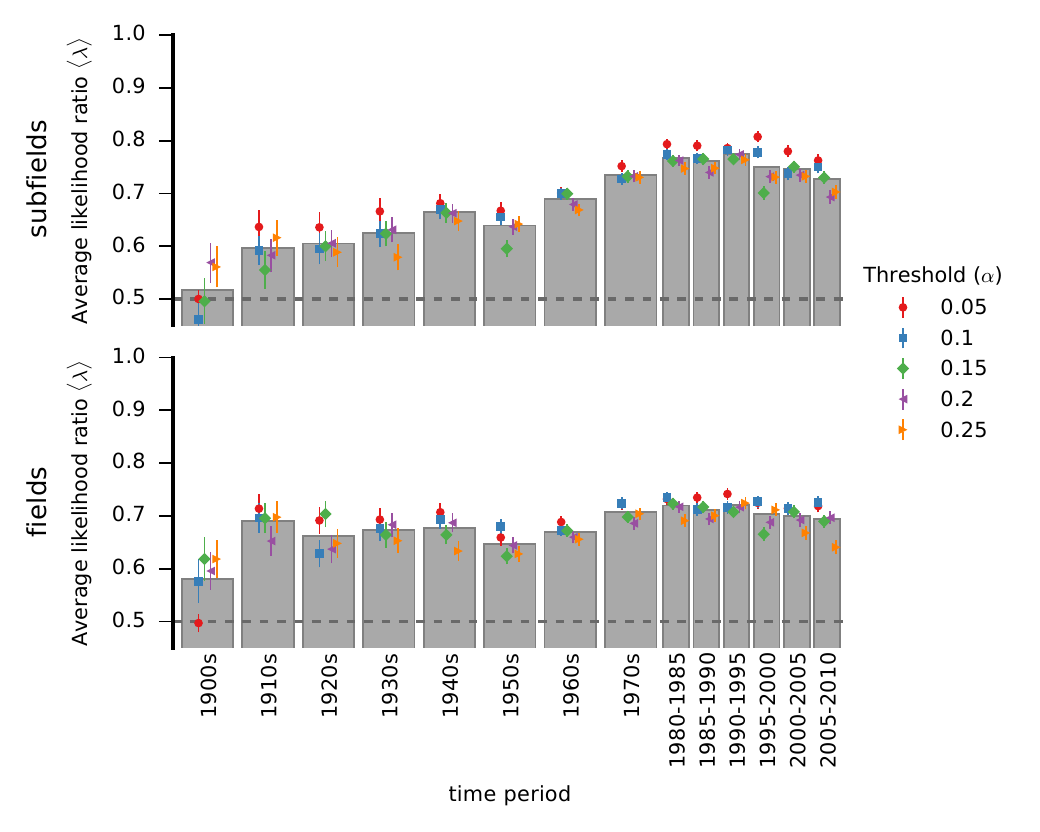}
      \caption{Node prediction performance, measured by the average predictive likelihood 
ratio $\avg{\lambda}$ for subfields and fields [see \protect\cref{eq:lambda}]. The $\avg{\lambda}$ values are 
calculated for simplified networks (see \cref{sec:data_simple}) for 14 time slices used previously, with five  
threshold values $\alpha$: $0.05,\ldots,0.25$. Each bar corresponds to a single time windows and the dots above 
each bar indicate the thresholds $\alpha$ with values increasing from left to right. 
Results are averaged over all the nodes 
from ten samples for each case, each one formed by removing 5\% or 100 nodes, whichever is 
smaller. The bar heights correspond to averages over
$\alpha$ values.}
  \label{fig:improvements}
\end{figure}

We use the average predictive likelihood ratio $\avg{\lambda}$ to measure the ability of subfield 
and field classifications to predict journals' citations in the simplified networks (see \cref{sec:data_simple}).
Predicting the links of the simplified networks is equivalent to predicting which journals are the most important sources and destinations for the 
citations of the extra nodes, because these networks only include the most important links for each node and do not include the actual
citation counts for the links.
 The values for $\avg{\lambda}$ are presented in \cref{fig:improvements} for each individual time 
slice, and for five threshold levels that preserve $\sim$6\% to $\sim$21\% of the most important links 
(representing $\sim$23\% to $\sim$45\% of citations) and $\sim$51\% to $\sim$99\% of 
nodes, respectively.

Overall, both subfield and field classification correlate positively with the citation 
structure. The only exception are the low threshold networks for 1900s, in which knowing 
the journal's (sub)field does not help in predicting its citations. Low threshold values in 
this already small network caused the loss of large fraction of links and nodes, which 
lowered the quality of the approximation by the simplified and thresholded network. 
Higher threshold values do not have this problem.

Subfields are more predictive with time while the predictiveness 
of the fields does not change as much. The reasons for this can be some sort of 
overspecialization of subfields, which do not 
necessarily correspond to the citation patterns of the journals being classified. Fields, 
on the other hand, remain as a good proxy for large-scale citation structure throughout 
the whole time period.

\subsection{Predictability of individual fields}
\label{sec:pred-field}

The method described in the previous Section answers the question of how much information 
we gain about journal's citations if we know what subfields it is classified into.
We will next divide this question into
smaller parts, and ask how much information do we gain by knowing that a journal belongs to a specific subfield.

Using the model from \citet{Hric2016} it is possible to calculate how much information gain (for 
guessing node's links) does a single annotation provide, in comparison to a case where 
annotations are assigned randomly. Information gain relative to the random case is 
defined as predictiveness 
$\mu_a$, and it is defined per annotation block $a$. Further 
details and formulas can be found in \citet{Hric2016}.

Here we again consider subfields and fields as annotations of journals.
After fitting the SBM onto the
two-layered network of citations and annotations,
the predictiveness 
of each block of subfields (or fields) is calculated.
Fields in the same block inherit predictiveness 
of the block, which follows from the SBM assumption that all annotations in a block are equivalent.
By calculating the field predictivenesses 
for all time windows, on top of being able to compare
fields to each other, their change in time can also be tracked both relatively and absolutely.

Because the implementation of SBM fitting algorithm is probabilistic, variations in results 
are to be expected with small differences in the networks (for instance two 10-year 
windows with 9-year overlap), and even in different runs of fitting function on the same 
network. These variations cause $\mu_a$ values for the network 
from adjacent time windows to vary considerably, obscuring more general trends.
Taking an average $\mu_a$ of all time windows that a year belongs to, we are able to
overcome these fluctuations. Additional clarity is achieved by ten-year sliding averages of these values.

\begin{figure}
      \centering
      \includegraphics[]{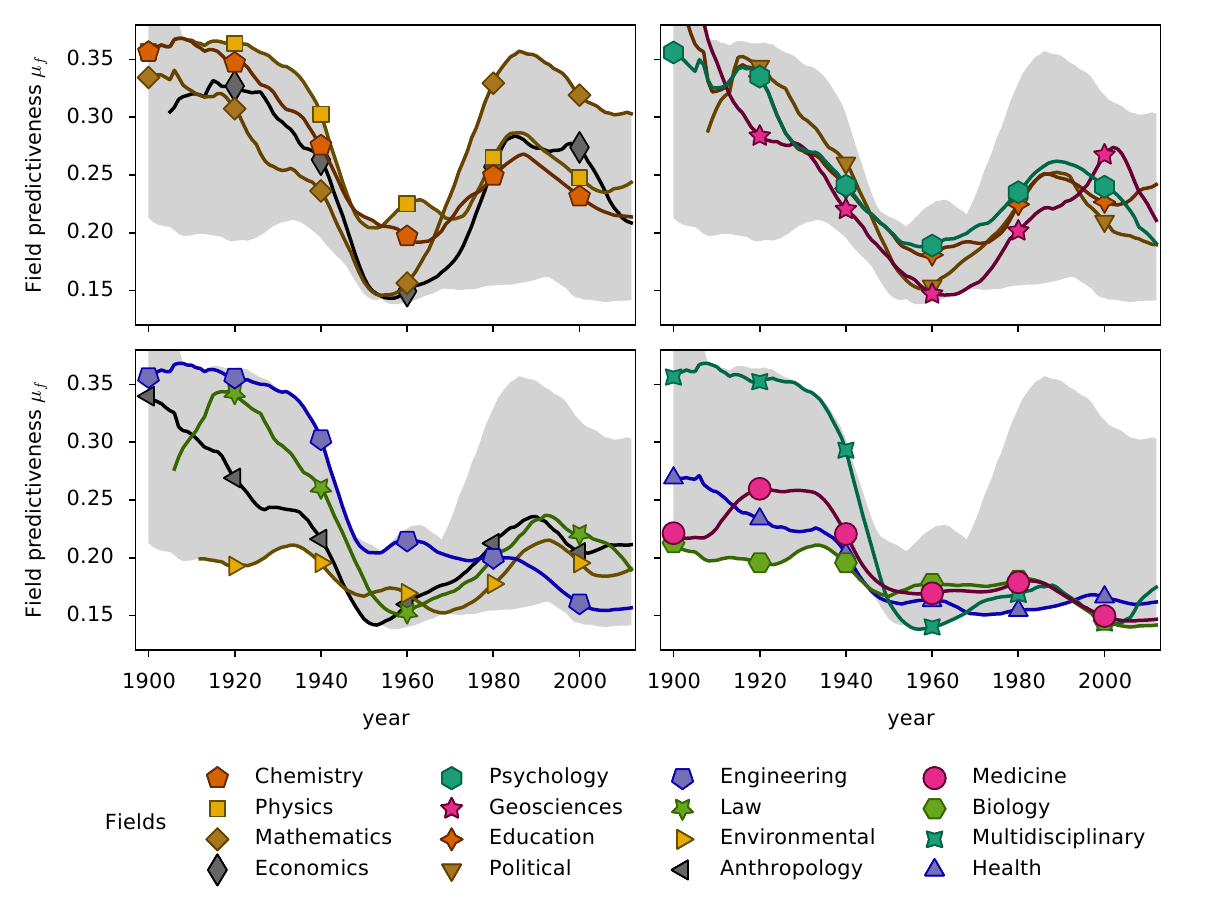}
  \caption{Predictivenesses measured by $\mu$ of the top 16 fields, for classification 
into fields, for all available years. The values are ten-year sliding averages of average
$\mu$ for all time windows using that year.
Each panel contains a groups of four fields in the order of decreasing $\mu$ after 1960.
Shaded region in the background is the total span of values is drawn as 
}
  \label{fig:field_imps}
\end{figure}

Here we present the predictivenesses 
of individual fields, for the case where classification into fields was used.
The results for the case where subfields were used instead, are presented in \cref{app:sf_predictiveness}.

Based on field predictiveness 
in \cref{fig:field_imps}, the time can be split into three periods: before 1940s, the transition period, and after 1970s. 
Before 1940s the fields have on average higher predictiveness, 
but since the data is quite scarce for this period, one needs to be careful when drawing conclusions. In the
transition period all the fields have very poor predictiveness, experiencing a rebound after 1970 for all but handful of fields in the last panels (Engineering, Environmental, Medicine, Biology, Multidisciplinary, and Health). 
This can be a sign of major changes science has gone through after the WWII.

Mathematics has the highest predictiveness 
in the third period by a visible margin, while before 
1940s the best scoring fields are Engineering and Multidisciplinary. It has risen 
sharply from the bottom in the transition period to the top in just 25 years, i.e. from 1955 to 1980.
This means that citation patterns of Mathematics papers became more characteristic after 
1970s, which is picked up by SBM and Mathematics journals end up in a small number of 
exclusive blocks.
For Engineering it is the opposite: it fared very high before 1940s, did not suffer hard 
in the transition, but never recovered. The same can be said about Multidisciplinary 
field: it had even sharper drop and did not really recover.

On the other side of the spectrum are often large fields (Engineering, Medicine, Biology),
or related to a large field (Environmental, Anthropology, and Health).
Their large size means that they contain rich structure within themselves, 
which gets detected by the SBM as large number of blocks. Hence knowing just the field label tells little about the small blocks within the field.

\section{Conclusions and Discussion}
\label{sec:Conclusions}

The tools developed in network science are routinely used to analyse the citation networks, and 
in particular network clustering methods are often used to identify large-scale patterns in these networks.
Recently, there has been significant advances in stochastic block modelling methods within the network science literature, 
but large-scale analysis of citation patterns using these methods have been missing until now. The main difference between the SBM methods and the conventional graph clustering 
is that the SBM methods can be used to detect a variety of mesoscale structures---not just dense subgraphs. 
Therefore, the SBM methods lead to compelling advantages over traditional network clustering only for networks in which
the dominant structure is not a collection of dense communities that are sparsely connected to each other.
In this article we have shown that this is the case for citation networks by fitting the SBMs to them: 
the blocks that best explain these networks are only rarely even weak communities.
This observation is true across multiple resolution levels and through the whole history of modern science.

The idea of finding sets of journals with similar citation patterns and similar roles is not new,
but the SBM formalises it in a way that is novel to citation network analysis.
Nodes in a block do not necessarily share the same links to all other nodes, but they do share similar links to
all other blocks. Previously similar structures were found
by calculating similarity matrices between journals (or other elements) and applying data clustering methods that work on the similarities.
This type of multi-stage method  requires researcher to first choose a similarity measure and then a clustering method and the type of clusters found 
reflect the unique combination of these two choices. The SBM method is more transparent: there is a single easy-to-interpret model that is fitted to the data.
Model fitting has its own pitfalls and problems, 
but these difficulties have been solved for SBM
in the recent literature leading to a very robust and efficient fitting methods.
This point is demonstrated by the fact that the SBM method works consistently with minimal prepossessing 
for both relatively small networks representing citation patterns in the 1900s and in the recent large-scale networks.

The journals are classified manually into subfields and fields, and these classes can be viewed as human-curated blocks. 
 Such  classification block structures should correspond to the structure of the citation networks, and it is interesting to study how these two are related. 
 A typical---although often flawed \citep{Hric2014}---approach for doing this is to consider the classification blocks as ``ground truth'' and compare them to structural blocks using some subset of the many measures  designed to quantify similarity or dissimilarity of partitions.
Here, we have taken a more direct and well-grounded approach to ask the question of how much can the field classifications help in predicting the 
citation patterns. We used the SBMs to test this, and found that overall the subfield classifications have become more helpful.
The ability to predict citations based on categorization was further split for different time periods and for different fields revealing 
a rich variety of temporal patterns.

We have demonstrated that SBMs are very suitable for the type of structure found in the citation networks, 
and therefore the SBM methods have the potential of becoming one of the standard tools in analysing citation networks in the future. 
The work presented here only lays a basis for such future analysis, even though we have analysed modern science in its full length and width.
There are  several immediate new research avenues and directions that are still interesting and open. 
Obviously, case studies can be focused on subsets of journals, at various levels of details, and for any time period, not just the ones considered here. 
Further, one can study how the rich annotation data of scientific
articles is related to the structure of the citation networks. For example, the relationships of keywords, authors, institutional affiliations, and countries 
to the structure of the networks can be analysed. The SBM method used here partitions annotations---in addition to journals---into blocks of equivalence, and these annotation blocks 
deserve to receive detailed attention.
For instance, one can find blocks of authors, institutions or countries that have the same role in the structure of the citation networks.
Lastly, rich hierarchical structures returned by SBM are ripe for comparison to existing ontologies that is left for future work.

\section{Acknowledgments}
\label{sec:acknowledgments}

We thank Tiago P. Peixoto for providing technical help regarding \texttt{graph-tool} Python module,
and useful comments on the final manuscript.
This work was supported by the European Community’s H2020 Program under the scheme ``INFRAIA-1-2014-2015: Research Infrastructures'', Grant agreement No. 654024 ``SoBigData: Social Mining \& Big Data Ecosystem'' (http://www.sobigdata.eu).
D.H. and K.K. gratefully acknowledge MULTIPLEX, grant number 317532 of the European Commission.
We acknowledge the computational resources provided by the Aalto Science-IT project.

\begin{appendices}

\section{Dataset: basic statistics}
\label[appsec]{app:basic_stats}

The total number of journals, articles and citations in the dataset is 76418, 55199417 (of 
which 38212193 have citations flowing in or out) and 632340116, respectively.

\begin{figure}
  \begin{tabular}{cc}
    \begin{minipage}{.46\columnwidth}
      \includegraphics[trim=9 0 0 0, clip=true]{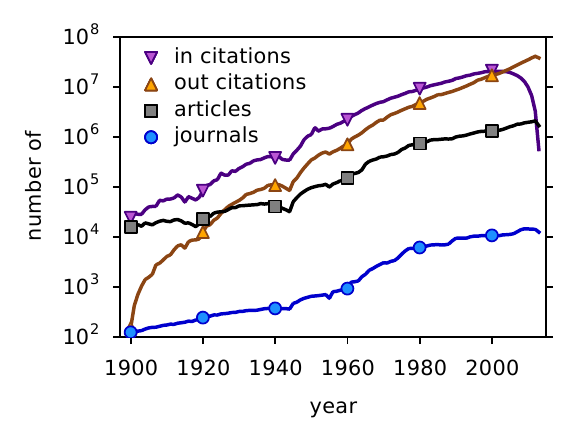}
    \end{minipage}
    \begin{minipage}{.49\columnwidth}
      \includegraphics[trim=8 0 0 0, clip=true]{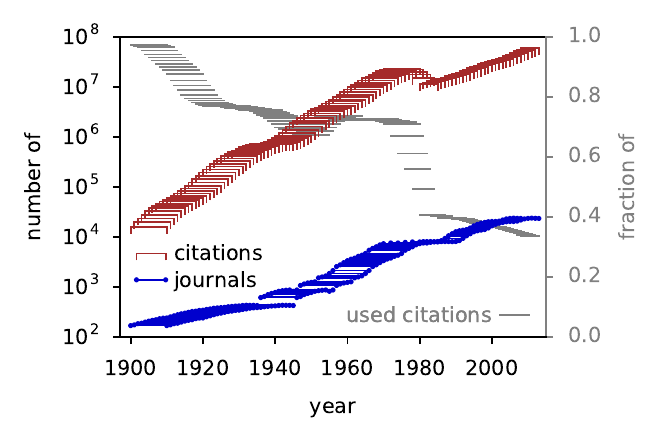}
    \end{minipage}
  \end{tabular}
  \caption{Number of journals, articles, and citations.
Left: values for every year in the dataset.
Right: values for every network made from slicing the data in time windows, including
the fraction of used citations out of all citations pertinent on journals in each time window.
}
  \label{fig:basics}
\end{figure}

The number of journals, articles, in- and out- citations for every year are presented in 
\cref{fig:basics}, left.
Exponential growth of science in all aspects is clearly visible. On the right are plotted numbers of 
journals and citations in networks for each time window. They are also growing 
exponentially which prompted us to use shorted time windows after 1970s. Due to using only 
the references not older than the length of the time window, only a certain fraction of citations is used,
which is plotted with grey lines on the right side of \cref{fig:basics}. Ten-year 
windows retain around 70\% of the citations, while 5-year ones around 35\%.

\begin{figure}
  \begin{tabular}{cc}
    \begin{minipage}{.52\columnwidth}
      \includegraphics[trim=8 0 0 0, clip=true]{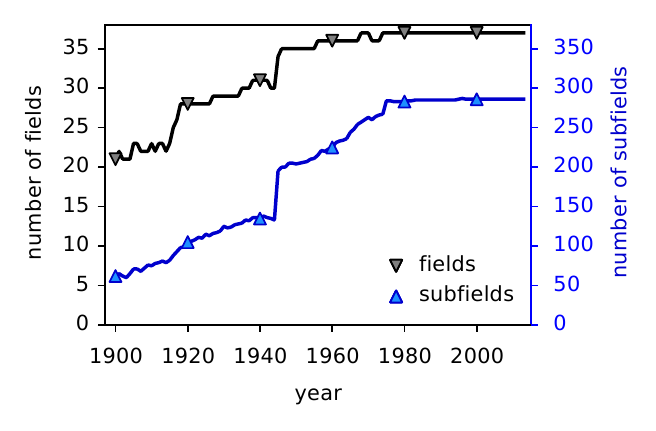}
    \end{minipage}
    \begin{minipage}{.49\columnwidth}
      \includegraphics[trim=3 0 0 0, clip=true]{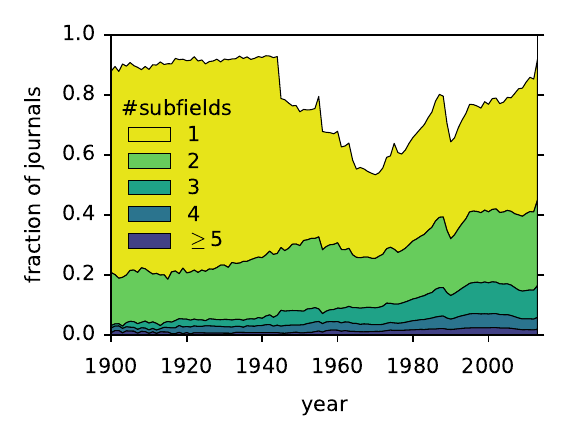}
    \end{minipage}
  \end{tabular}
  \caption{Number and distribution of subject categories.
Left: number of fields and subfields for every year.
Right: fractions of journals classified in different number of subfields.
}
  \label{fig:basics_tags}
\end{figure}

In contrast to all other measures, number of subject categories grows only linearly in 
time (\cref{fig:basics_tags}, left). It needs to be pointed out that these values do 
not necessarily represent the reality truthfully, since
it is assumed that classifications of a journal remain the 
same throughout its lifespan, for simplicity reasons. Fractions of classified journals depicted on the right side 
of \cref{fig:basics_tags} show that most journals are classified to at least one subfield, 
with the lowest fraction of classified journals being around year 1970.

\section{Predictiveness of individual fields}
\label[appsec]{app:sf_predictiveness}

\begin{figure}
      \centering
      \includegraphics[]{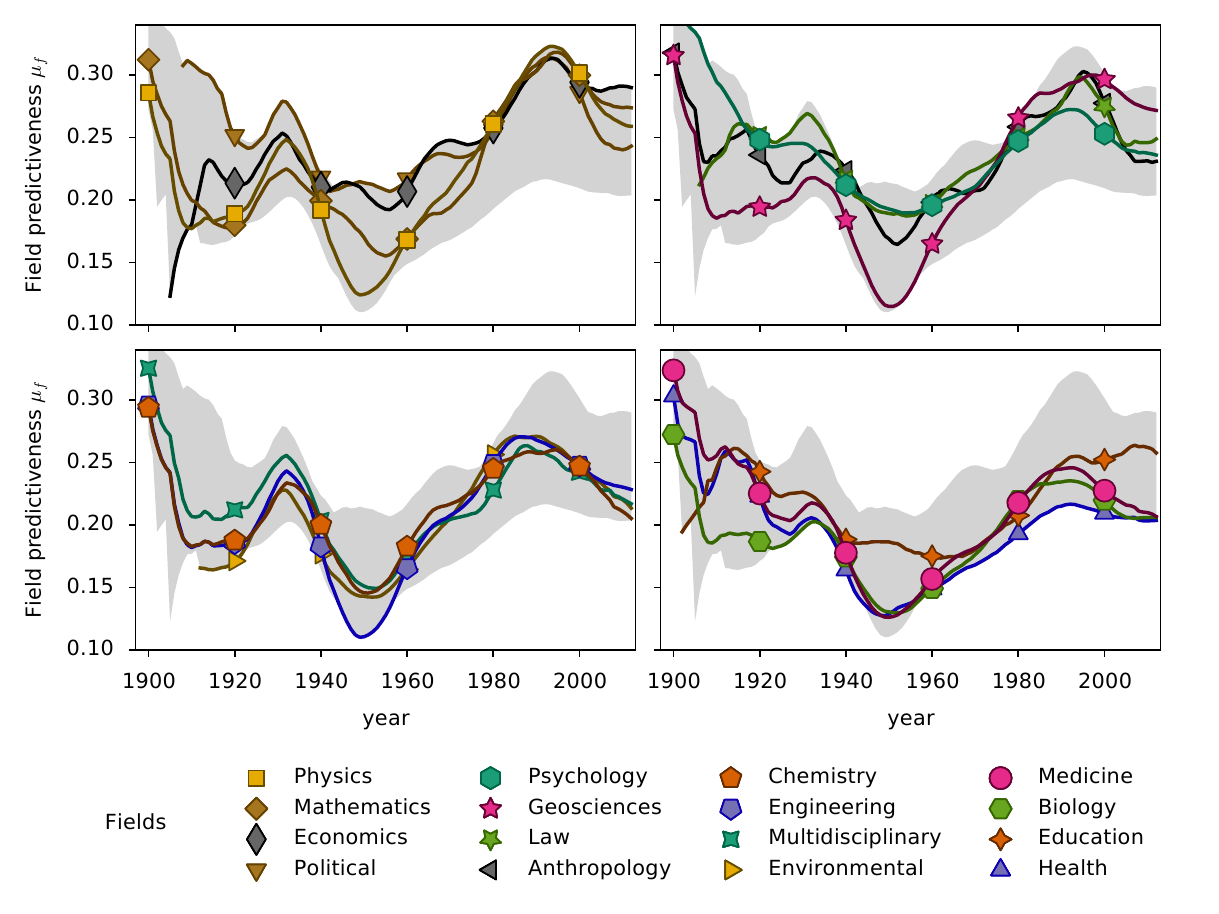}
  \caption{Predictivenesses measured by $\mu$ of the top 16 fields, for classification 
into subfields, for all available years. The values are the averages of overlapping time windows,
additionally smoothed by 10-year sliding average.
Each panel contains a groups of four fields in the order of decreasing $\mu$ after 1960.
Total span of values is drawn as shaded region in the background.
}
  \label{fig:field_imps_sf}
\end{figure}

Field predictivenesses ($\mu_{f}$) from \cref{sec:pred-field} are calculated using field categories as 
``meta-information''. The same can be calculated for the case when subfields are used 
instead, the results of which are presented in \cref{fig:field_imps_sf}. Although 
$\mu_{sf}$ values are calculated for subfields, values for fields are calculated as averages of all subfields belonging to a 
field. In addition to averages, errors of the mean are shown as shaded regions.

In general, subfields are slightly better predictors of journal's citations than fields. 
This is true for all subfields, regardless of the field they belong to, which in 
consequence prevents clear separation of fields based on this criteria. Nevertheless,
the order of fields by predictiveness of their subfields is roughly the same as for the
predictiveness of fields themselves (c.f. \cref{fig:field_imps} in \cref{sec:pred-field}).

\section{Formulas for connectivity patterns of merged blocks}
\label{app:block_merge}

\begin{figure}
      \centering
      \includegraphics[]{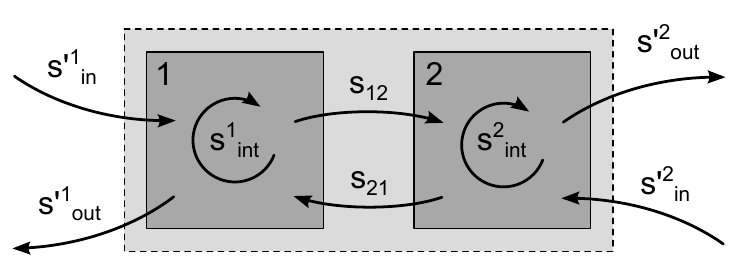}
  \caption{Schematic representation of citation flows in the merge of two blocks.
}
  \label{fig:block_merge}
\end{figure}

Each level in a hierarchy takes the blocks from the level below it as nodes of the network that it models.
In this process, the properties of the small blocks ($\bar{s}_{int}$, $\bar{s}_{in}$, and $\bar{s}_{out}$) are propagated in non-trivial way into the properties of the merged block.

Here we consider the properties of a block that is formed when two blocks at the lower level are merged (see \cref{fig:block_merge}).
Internal flow of the merged block ($S_{int}$) contains all flows inside a dashed rectangle on \cref{fig:block_merge}, while in- and out- flows ($S_{in}$, $S_{out}$) are sums of flows of the constituent blocks:
\begin{equation}
  \begin{aligned}
    S_{int} &= s^1_{int}+s^2_{int}+s_{12}+s_{21}, \\
    S_{in}  &= s'^1_{in}+s'^2_{in}, \\
    S_{out} &= s'^1_{out}+s'^2_{out}.
  \end{aligned}
  \label{eq:S_def}
\end{equation}
Normalized flows of the merged block are obtained by dividing the absolute flows by the total flows of the block exactly as also described in \cref{eq:s_def}:
\begin{equation}
  \begin{aligned}
    \bar{S}_{int} &= S_{int}/S_{tot}, \\
    \bar{S}_{in}  &= S_{in}/S_{tot}, \\
    \bar{S}_{out} &= S_{out}/S_{tot}.
  \end{aligned}
  \qquad \text{where} \quad S_{tot}=S_{int}+S_{in}+S_{out}.
  \label{eq:Sbar_def}
\end{equation}
Expressing the normalized flows of the merged block with the normalized flows of the constituent blocks, we get:
\begin{equation}
  \begin{aligned}
    \bar{S}_{int} &= \frac{s^1_{tot}}{S_{tot}}\bar{s}^1_{int}
                   + \frac{s^2_{tot}}{S_{tot}}\bar{s}^2_{int}
                   + \frac{s_{12}+s_{21}}{S_{tot}}, \\
    \bar{S}_{in}  &= \frac{s^1_{tot}}{S_{tot}}\bar{s}^1_{in}
                   + \frac{s^2_{tot}}{S_{tot}}\bar{s}^2_{in}
                   - \frac{s_{12}+s_{21}}{S_{tot}}, \\
    \bar{S}_{out} &= \frac{s^1_{tot}}{S_{tot}}\bar{s}^1_{out}
                   + \frac{s^2_{tot}}{S_{tot}}\bar{s}^2_{out}
                   - \frac{s_{12}+s_{21}}{S_{tot}}, \\
  \end{aligned}
  \label{eq:Smerged_def}
\end{equation}
where
\begin{equation}
  s^b_{tot}=s^b_{int}+s^b_{in}+s^b_{out}+s_{12}+s_{21}, \quad \text{for} \; b \in \{1,2\}.
\end{equation}
The three formulas from \cref{eq:Smerged_def} have the same form:
\begin{equation}
  \bar{S}_{f} = \frac{s^1_{tot}}{S_{tot}}\bar{s}^1_{f}
              + \frac{s^2_{tot}}{S_{tot}}\bar{s}^2_{f}
              \pm \frac{s_{12}+s_{21}}{S_{tot}}
              = w_1 \bar{s}^1_f + w_2 \bar{s}^2_f \pm w_{12}.
  \label{eq:Sf}
\end{equation}
where $w_1$, $w_2$, are the fractions of flows pertinent to blocks 1 and 2,
and $w_{12}$ is the fraction of flow that goes between blocks 1 and 2,
out of total flow $S_{tot}$.

\end{appendices}

\bibliography{bib}
\bibliographystyle{spbasic-nolinks}

\end{document}